\documentclass[conference,compsoc]{IEEEtran}
\usepackage{graphicx}
\usepackage{booktabs}
\usepackage{multirow} 
\usepackage{array}
\usepackage{makecell}
\usepackage{tabularx}
\usepackage{booktabs,tabularx,siunitx}
\usepackage{threeparttable}
\usepackage{xcolor}
\usepackage[most]{tcolorbox}
\usepackage{url}
\usepackage{threeparttable}

\usepackage{enumitem}

\ifCLASSOPTIONcompsoc
  \usepackage[nocompress]{cite}
\else
  \usepackage{cite}
\fi

\ifCLASSINFOpdf
\else
\fi
\begin{document}
\title{“What I Sign Is Not What I See”:

Towards Explainable and Trustworthy Cryptocurrency Wallet Signatures}


% \author{\IEEEauthorblockN{Michael Shell}
% \IEEEauthorblockA{School of Electrical and\\Computer Engineering\\
% Georgia Institute of Technology\\
% Atlanta, Georgia 30332--0250\\
% Email: http://www.michaelshell.org/contact.html}
% \and
% \IEEEauthorblockN{Homer Simpson}
% \IEEEauthorblockA{Twentieth Century Fox\\
% Email: homer@thesimpsons.com}
% \and
% \IEEEauthorblockN{James Kirk\\ and Montgomery Scott}
% \IEEEauthorblockA{Starfleet Academy\\
% San Francisco, California 96678-2391\\
% Telephone: (800) 555--1212\\

\author{
\IEEEauthorblockN{Yuyang Qin}
\IEEEauthorblockA{
The Chinese University of Hong Kong, Shenzhen\\
yuyangqin1@link.cuhk.edu.cn
}
\and
\IEEEauthorblockN{Haihan Duan\textsuperscript{*}}
\IEEEauthorblockA{
Shenzhen MSU-BIT University\\
duanhaihan@smbu.edu.cn
}
}


% compsoc conferences compared to traditional conferences), use this
%Telephone: (800) 555--1212, Fax: (888) 555--1212}


\maketitle

\begin{abstract}
Cryptocurrency wallets have become the primary gateway to decentralized applications, yet users often face significant difficulty in discerning what a wallet signature actually does or entails. Prior work has mainly focused on mitigating protocol vulnerabilities, with limited attention to how users perceive and interpret what they are authorizing. To examine this usability–security gap, we conducted two formative studies investigating how users interpret authentic signing requests and what cues they rely on to assess risk. Findings reveal that users often misread critical parameters, underestimate high-risk signatures, and rely on superficial familiarity rather than understanding transaction intent. Building on these insights, we designed the Signature Semantic Decoder---a prototype framework that reconstructs and visualizes the intent behind wallet signatures prior to confirmation. Through structured parsing and semantic labeling, it demonstrates how signing data can be transformed into plain-language explanations with contextual risk cues. In a between-subjects user study (N = 128), participants using the prototype achieved higher accuracy in identifying risky signatures, improved clarity and decision confidence, and lower cognitive workload compared with the baseline wallet interface. Our study reframes wallet signing as a problem of interpretability within secure interaction design and offers design implications for more transparent and trustworthy cryptocurrency wallet interfaces.
\end{abstract}


% \fi
\IEEEpeerreviewmaketitle

\section{Introduction}
Cryptocurrency wallets (hereinafter referred to as the ``wallet") constitute the primary interface through which users interact with decentralized applications (dApps)~\cite{Frohlich2022BlockchainHCI,Elsden2018MakingSense,UserCenteredDApps2020}. A critical step in these interactions is \textit{signing}, where users authorize cryptographic messages to execute on-chain or off-chain actions. Unlike conventional confirmations such as password entry or explicit consent dialogs, wallet signatures are typically presented as opaque hexadecimal strings or structured JSON payloads~\cite{voskobojnikov2021u,HolisticWallet2021NDSS}, making it nearly impossible for general users to comprehend what they are approving. This lack of semantic transparency creates a profound usability–security gap with direct implications for user safety security.

The opacity of current signing interfaces has facilitated a wide range of attacks, including phishing, blind signing, and social engineering, where users unknowingly approve malicious smart contracts or grant excessive permissions~\cite{Geels2024Signatures,BlockchainWalletUsability2020,SecurityAspectsWallets2023}. Existing mitigations, such as static warnings or minimal field labeling, operate at the interface rather than the semantic level, providing users with partial cues but not meaningful understanding~\cite{Zhou2024CrudiTEE,Kim2023HBET}. From a human–computer interaction (HCI) perspective, users are effectively asked to make trust decisions about cryptographic payloads that possess technical integrity yet lack human interpretability~\cite{Frohlich2022BlockchainHCI,Elsden2018MakingSense,UserCenteredDApps2020}. Although usable security research has long examined key management, authentication, and warning design~\cite{Kwon2023UXRegression,SecurityAspectsWallets2023}, the interpretability of blockchain signatures remains largely uncharted. Current wallets (e.g., MetaMask~\cite{MetaMask} and Rabby Wallet~\cite{RabbyWallet}) offer limited risk indicators and fail to reconstruct transaction intent into a human-centered narrative, leaving users vulnerable to deception and cognitive overload during the signing process~\cite{voskobojnikov2021u,HolisticWallet2021NDSS,BlockchainWalletUsability2020,Kim2023HBET}.

To address this gap, our study investigates how \textit{semantic transparency}—the explicit decoding and contextual explanation of signing data—affects users’ comprehension, trust, and decision-making in cryptocurrency wallet interactions. Specifically, we focus on three research questions:

 \begin{itemize}
     \item \textbf{RQ1:} To what extent does semantic transparency improve users’ ability to identify deceptive or risky signatures without increasing cognitive workload?
     \item \textbf{RQ2:} How does semantic transparency affect users’ comprehension, confidence, and perceived control when evaluating wallet signature requests?
     \item \textbf{RQ3:} How do users perceive the role of semantic transparency in shaping their trust and overall signing experience?
 \end{itemize}

To systematically address these questions, we adopted a multi-stage research process comprising two formative studies, a system design phase, and a controlled main study. The first formative study examined how users interpret raw wallet signature data and revealed key sources of misunderstanding and misplaced confidence in existing signing interfaces. The second formative study investigated users’ perception of semantic cues and risk indicators through simulated signing scenarios, deriving design requirements for an interpretability-centered wallet interface. Building on these insights, we designed and implemented the \textit{Signature Semantic Decoder}—a middleware framework that reconstructs the meaning of signing payloads through structured parsing, semantic labeling, action inference, and contextual visualization. At last, we conducted a controlled experiment comparing the semantic interface against a baseline wallet, evaluating users’ comprehension, workload, risk awareness, and trust through both quantitative and qualitative analyses.

Taken together, this work advances the understanding of human–wallet interaction by showing how semantic transparency enhances both usability and security in decentralized signing, through three key contributions:

\textbf{First}, this research bridges the gap between opaque, machine-oriented signing interfaces and users’ need for interpretable, trustworthy cues. By introducing semantic transparency as a design principle, it establishes a systematic framework for reconstructing the meaning of blockchain signatures, helping users understand what they approve in a human-centered and contextualized way.

\textbf{Second}, the findings offer empirical insights into how semantic explanations shape users’ comprehension, workload, and risk perception in high-stakes cryptographic decisions. The results demonstrate that semantic decoding substantially improves users’ ability to detect deceptive or risky signatures and fosters informed, confident signing behavior without increasing cognitive burden.

\textbf{Third}, the evaluated approach contributes a reusable methodological and technical foundation for designing interpretable wallet infrastructures. We also make our prototype implementation and associated artifacts publicly available to facilitate further research\footnote{The code will be made publicly available upon publication.}. Beyond the proposed prototype, the insights can inform future wallet architectures, regulatory-compliant signing standards, and broader explainable security mechanisms where interpretability functions as both a usability enhancement and a security control.

\section{BACKGROUND AND RELATED WORK}

\subsection{Wallets and Web3 Signatures}
Cryptocurrency wallets serve as the fundamental interface through which users manage digital assets and interact with decentralized applications (dApps)~\cite{Frohlich2022BlockchainHCI,Elsden2018MakingSense}. Functionally, wallets operate as cryptographic key managers that generate, store, and use private keys to authorize user actions. Every transaction, smart contract call, or off-chain authentication is mediated through a digital signature, which guarantees authenticity and integrity in blockchain execution models~\cite{WoodYellowPaper}. Without the private key and its corresponding signature, ownership and authorization in blockchain systems cannot be verified~\cite{WoodYellowPaper}.

Most mainstream platforms such as Ethereum employ the Elliptic Curve Digital Signature Algorithm (ECDSA) over the \textit{secp256k1} curve~\cite{SEC1}. While cryptographically sound, this architecture optimizes for machine-level verifiability rather than human-level comprehension. Signatures are readily validated by programs but remain opaque to humans, typically presented as hexadecimal strings or verbose JSON payloads~\cite{voskobojnikov2021u,voskobojnikov2021u}. Wallet signing was thus engineered for computational verification, not human interpretation, which forces users to approve security-critical operations they rarely understand~\cite{SecurityAspectsWallets2023}.

In practice, the signing process typically follows a standardized workflow: decentralized applications construct a payload that encodes transaction parameters or message content, and the wallet presents this data to the user for authorization before generating a digital signature. Most mainstream wallets provide limited interpretive aids, such as displaying token symbols, recipient addresses, or transaction amounts. However, these cues operate at the surface level and rarely convey the underlying intent or consequence of the action~\cite{voskobojnikov2021u,voskobojnikov2021u,Kim2023HBET}. As a result, users often see partial fragments of meaning rather than a coherent narrative of what their signature authorizes. When contextual information is incomplete or ambiguous, seemingly routine confirmations can conceal high-risk operations such as unlimited approvals or off-chain delegations~\cite{Yan2024BlindMessage,SecurityAspectsWallets2023}. The signing interface therefore remains a cognitively demanding space, which verifies authenticity with cryptographic precision yet offers only limited semantic transparency to human users.

This opacity has direct security implications. Blind signing—approving opaque signatures without understanding—has become a major attack vector for phishing, approval fraud, and permission escalation in decentralized finance~\cite{Yan2024BlindMessage,SecurityAspectsWallets2023}. Since blockchain transactions are irreversible, a single uninformed signature can irretrievably transfer assets or grant malicious access~\cite{yu2024walletchoice}. Modern wallet signatures reproduce a similar paradox identified by Whitten and Tygar in their seminal work \textit{Why Johnny Can’t Encrypt}~\cite{Whitten1999Johnny}: they are mathematically secure yet practically opaque. Accordingly, our study shifts the focus from the syntactic design of signatures to their \textit{semantic interpretability} by investigating whether users can comprehend the real-world meaning and intent of what they authorize.

\subsection{Signing Methods and Security Risks}

Cryptocurrency wallets primarily rely on a small set of widely adopted signing methods standardized through Ethereum Improvement Proposals (EIPs)~\cite{EIP191,EIP712}, which cover the majority of contemporary wallet–dApp interactions~\cite{Meisami2025SigScope} and are summarized in Table~\ref{tab:signature_comparison}. These methods form the bridge between user intent and blockchain execution, encoding a wide range of actions into signed payloads, from token transfers to off-chain authentications. Although they share a unified cryptographic foundation, their structural and semantic characteristics differ substantially, shaping both usability and security outcomes~\cite{SecurityAspectsWallets2023}.

\begin{table}[h]
\caption{Comparison of Different Signing Methods}
\label{tab:signature_comparison}
\renewcommand{\arraystretch}{1.08}
\setlength{\tabcolsep}{2.5pt}
\begin{tabularx}{\columnwidth}{@{}l c c c c@{}}
\toprule
\textbf{Category} & \textbf{Method} & \textbf{Readability} & \textbf{Prefix} & \textbf{Data Structure} \\
\midrule
On-chain & \texttt{Tx-sign} & Partial & Implicit & Transaction object \\
\midrule
\multirow{3}{*}{Off-chain}
& \texttt{personal\_sign} & Low & Yes & UTF-8 message \\
& \texttt{eth\_sign} & Low & No & Arbitrary bytes \\
& \texttt{signTypedData} & Medium & Yes & Structured object \\
\bottomrule
\end{tabularx}
\vspace{2pt}

{\footnotesize \textit{Note:} \texttt{eth\_sign} has been deprecated and should not be used in any dApp or wallet due to its high security risks~\cite{SecurityAspectsWallets2023}}.

\end{table}

\textbf{On-chain signing.}  
The canonical on-chain mechanism, \textit{eth\_sendTransaction}, directly authorizes transactions that are executed and recorded on the blockchain. Its strength lies in transparency and verifiability: all parameters, such as sender, recipient, value, gas, and contract function, are part of the immutable transaction record, allowing deterministic verification by nodes~\cite{WoodYellowPaper}
. However, this low-level explicitness does not naturally translate into human comprehension. Although modern wallets render some fields in more readable forms, they often still present raw addresses, ABI-encoded function selectors, or hexadecimal parameter payloads when handling complex contract interactions, leaving users with limited insight into the actual semantics of the request~\cite{panicker2024end,wang2022penny}.  As a result, malicious operations such as hidden token approvals or unauthorized state changes can be disguised as routine transactions~\cite{voskobojnikov2021u,yu2024walletchoice}. Because blockchain transactions are irreversible, a single misunderstanding can lead to serious financial consequences.

\textbf{Off-chain signing.}  
In contrast, off-chain signing methods (\textit{eth\_sign}, \textit{personal\_sign}, and the \textit{eth\_signTypedData} family) authorize data that are verified cryptographically but not immediately executed on-chain. This indirection brings important advantages: it enables authentication, delegation, or pre-authorization without incurring transaction fees, and allows developers to design flexible application-layer protocols~\cite{Yan2024BlindMessage,voskobojnikov2021u}. Yet these benefits come with weakened contextual binding and reduced interpretability. Because the signed content is often detached from on-chain state, users cannot easily trace its eventual effect or scope of permission~\cite{SecurityAspectsWallets2023}. Simpler methods such as \textit{eth\_sign} or \textit{personal\_sign} provide minimal structure, making phishing and replay attacks feasible, while structured standards like EIP-712 improve verifiability but still rely on JSON schemas that overwhelm users with technical details~\cite{Kim2023HBET}.

Among them, \textit{eth\_sign} represents the lowest level of abstraction, signing arbitrary byte sequences without message prefixes or domain separation. This flexibility allows signatures to be reused across contexts, a property attackers exploit to perform replay or phishing attacks, leading to its deprecation in modern dApps~\cite{SecurityAspectsWallets2023}.  
\textit{personal\_sign}, introduced in EIP-191~\cite{EIP191}, improved contextual safety through a fixed prefix—``\texttt{\textbackslash x19Ethereum Signed Message:}''—that binds the message to the Ethereum domain. However, since it lacks structured semantics, the message body itself may still conceal authorization payloads or deceptive prompts (e.g., “verify your account to continue”), making the interface only superficially transparent.  
EIP-712~\cite{EIP712} (\textit{signTypedData}) extends this model by introducing typed, schema-defined structures that specify message domains and field hierarchies. This design improves machine-verifiable intent but does not necessarily enhance human interpretability: wallets typically render raw JSON schemas that overwhelm users with technical parameters, obscuring the real implications of their approval~\cite{voskobojnikov2021u,yu2024walletchoice}.  

Across these mechanisms, users face uneven interpretability across signing formats. Simpler methods such as \textit{personal\_sign} reveal limited cues (like readable text or familiar addresses) that support only surface-level judgment, whereas structured standards such as \textit{signTypedData} (EIP-712) provide richer schemas that enhance technical verifiability but still fall short of conveying intent in human terms~\cite{voskobojnikov2021u,Kim2023HBET}. As a result, even when key fields appear legible, users often fail to grasp the broader logic or implications of unfamiliar contracts. In practice, wallets present a trade-off between flexibility and transparency—where interoperability often comes at the cost of interpretability~\cite{watanabe2024vellet,voskobojnikov2021u}. This structural asymmetry underlies many real-world threats such as blind signing, phishing, and unlimited approval scams~\cite{Yan2024BlindMessage,SecurityAspectsWallets2023}. Addressing these vulnerabilities therefore requires not only stronger technical safeguards but also a shift toward semantic transparency, enabling users to comprehend the intent and consequence of what they sign.

\subsection{Wallet Usability and Prior Studies}

Usable security research has long shown that technically robust mechanisms often fail when they overlook human cognition~\cite{Whitten1999Johnny,Sunshine2009CryingWolf,asgharpour2007mental}.
 Early work argued that users are not the enemy but victims of poor security design~\cite{Adams1999Users}, while the notion of a “compliance budget’’ emphasized that security decisions compete with limited attention and cognitive resources~\cite{Beautement2009Compliance}. Subsequent studies further demonstrated that exposing users directly to cryptographic primitives such as keys or hashes leads not only to confusion and errors but also to insecure workarounds that undermine intended protections~\cite{Krol2016Keyring,Redmiles2020SecurityReview}. Collectively, this line of work highlights a persistent lesson: security mechanisms that ignore usability ultimately weaken their own guarantees.

In blockchain systems, this tension is amplified. Wallets function simultaneously as key managers and as decision interfaces through which users authorize financial or contractual actions, giving the signing moment disproportionate weight. Early work on key recovery and seed-phrase handling documented anxiety around irreversible loss, fragile storage practices, and the absence of support during recovery events~\cite{Eskandari2018Bitcoin,Abramova2021Bits}. More recent studies examined transaction comprehension, finding that users often misread trust boundaries, assume reversibility, and overlook critical cues during confirmation screens~\cite{voskobojnikov2021u,Frohlich2022BlockchainHCI,Kim2023HBET}. Yet despite these insights, the interpretability of the signing process itself—when users must connect low-level cryptographic data to real-world intent—remains comparatively underexplored.

The act of signing is therefore the most decisive yet least understood step in wallet workflows. Many users “click through’’ signing prompts, treating them as routine confirmations rather than high-stakes authorizations~\cite{voskobojnikov2021u,yu2024walletchoice}. This \textit{blind signing} emerges from a fundamental mismatch between cryptographic representation and user understanding: interfaces display technically valid but semantically opaque payloads that reveal little about purpose or consequence. Mainstream wallets such as MetaMask and Phantom~\cite{PhantomWallet} leave this interpretive burden to users, while newer tools like Rabby or GoPlus~\cite{GoPlusSecurityService} add contextual warnings and heuristics but still fall short of reconstructing meaning in a systematic way~\cite{SecurityAspectsWallets2023,Yan2024BlindMessage}. These reactive strategies address surface symptoms rather than underlying causes.

At the same time, system-level research has examined structural and implementation risks in wallet signing. \textit{SigScope}~\cite{Meisami2025SigScope}, for example, conducts large-scale static and dynamic analysis across thousands of dApps, revealing widespread vulnerabilities tied to inconsistent or unsafe signature handling, while He et al.~\cite{he2020security} similarly identifies recurring weaknesses in key protection, transaction validation, and verification logic. Recent work by Yan et al.~\cite{yan2024stealing} systematically analyzes Web3 authentication flows built on \textit{}t{personal\_sign}, identifying blind-message, replay, and multi-message attacks in which missing origin fields and insufficient server-side verification allow attackers to relay signed login messages across sites and gain unauthorized access.
 Although these studies significantly advance protocol and implementation security, they leave the \textit{human interpretability} of signature content largely untouched. Even experienced users rely on social heuristics—such as interface familiarity, brand trust, or habitual clicking—when authorizing transactions~\cite{yu2024walletchoice}, while unstructured methods like \textit{personal\_sign} continue to facilitate phishing and session-binding attacks that exploit semantic opacity~\cite{Yan2024BlindMessage}. Existing wallets therefore mitigate known risks reactively but do not provide a coherent understanding of what a signature entails. This gap motivates our focus on \textit{semantic transparency}: the extent to which wallets can decode, explain, and visualize signing intent in human-understandable terms, enabling more interpretable and trustworthy signing experiences.

\section{Formative Studies and System Design}

To ground the development of the \textit{Signature Semantic Decoder} in empirical evidence, we conducted two formative studies examining how users perceive and evaluate crypto wallet signature requests in realistic contexts. The findings revealed recurring patterns of ambiguity, over-trust, and limited risk awareness, which informed the subsequent system design. The following sections present the formative results (Sections~\ref{sec:formative1}--\ref{sec:formative2}) and outline the architecture, workflow, and interface of the proposed system (Sections~\ref{designgoal}--\ref{ui}).

\subsection{Formative Study 1: User Experience}
\label{sec:formative1}

\begin{figure}[t]
  \centering
  \vspace*{-0.7cm}
  \hfill
  \includegraphics[width=\columnwidth]{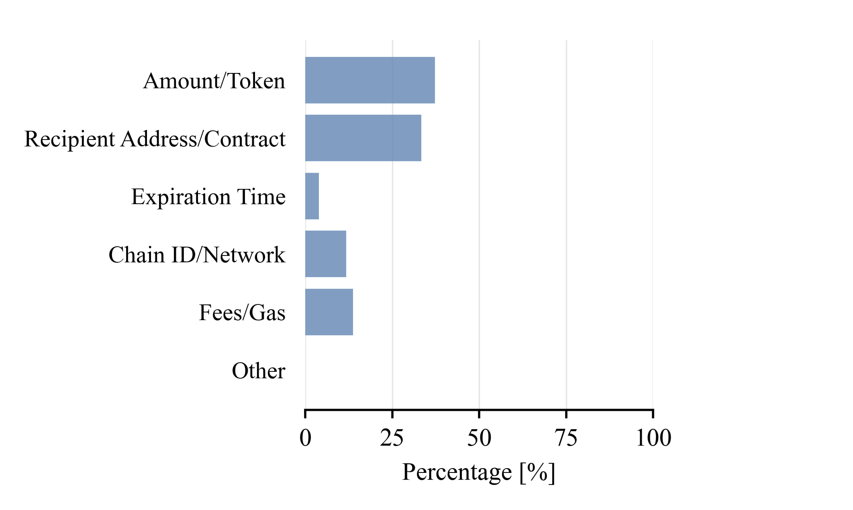}
  \caption{User attention distribution in signature requests.}
  \label{pilot1.1}
\end{figure}

This formative study investigated how users perceive, interpret, and act upon wallet signature requests in Web3 contexts. Based on responses from 50 participants across twenty survey questions, we found that while most respondents were experienced wallet users—over half had engaged in DeFi or NFT activities and 75\% had used wallets for more than a year, their comprehension of signing data remained limited. As shown in Figure~\ref{pilot1.1}, participants tended to focus on surface-level transactional cues, with 78\% checking the token amount and 71\% the recipient address, whereas parameters such as gas fee, chain ID, or expiration time were rarely examined. This attention pattern suggests that even seasoned users approach signing as a familiar financial confirmation rather than an informed authorization act, prioritizing visible monetary information over technical or contextual fields that define the real execution scope.

\begin{figure}[t]
  \centering
  \includegraphics[width=\columnwidth]{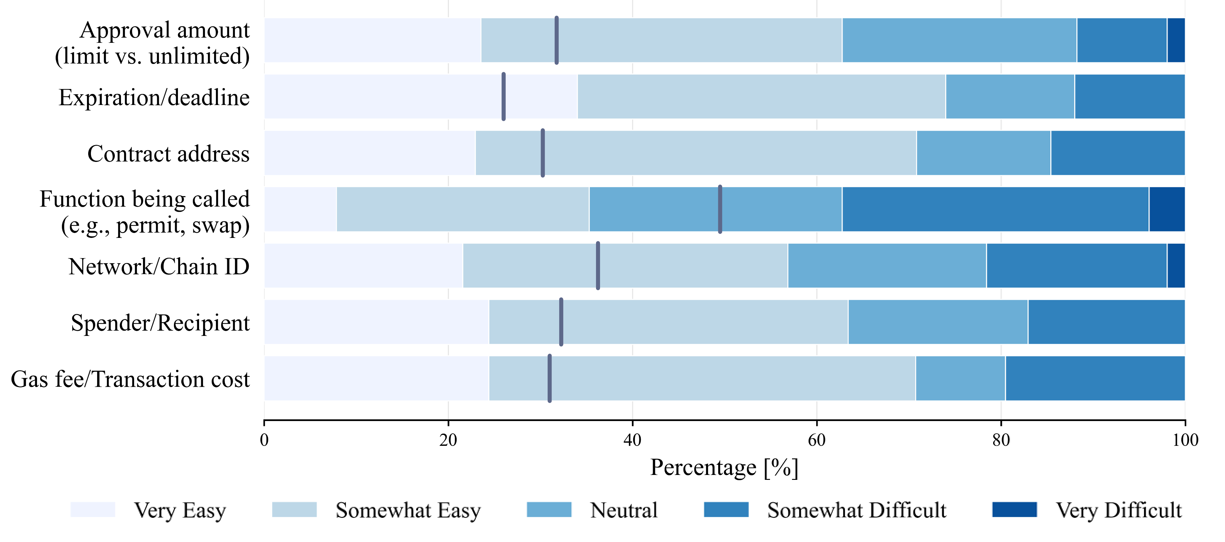}
  \caption{Perceived difficulty of understanding different signature elements.}
  \label{pilot1.2}
  \vspace*{-0.3cm}
\end{figure}

When asked about specific elements within a signature request (Figure~\ref{pilot1.2}, users rated approval limits and expiration deadlines as relatively easy to understand (mean = 2.27 and 2.04), but struggled significantly with function names (mean = 2.98) and network identifiers (mean = 2.45). Many described contract calls such as \textit{permit} or \textit{swapExactTokens} as ``opaque'' or ``developer-only,'' revealing a semantic gap between system language and user cognition. Complementary questions further showed that 64\% of users primarily trust familiar wallet interfaces or recognizable dApp brands, while only 32\% relied on verifiable contract data. Together, these results reveal that user trust and understanding remain interface-driven rather than data-driven: familiarity substitutes for clarity, and perceived safety derives from visual consistency rather than semantic transparency. Together, these findings indicate that users’ trust formation is anchored in interface familiarity rather than verifiable data, revealing a structural misalignment between the surface cues provided by wallets and the underlying semantics of authorization.

\subsection{Formative Study 2: User Decision-Making}
\label{sec:formative2}

To deepen our understanding of how users interpret and respond to signature requests, we conducted a follow-up study with eight representative tasks (T1 to T8) that covered login, token approval, bridging, NFT minting, and governance voting. Of 30 participants most had used wallets for more than one year (77\%) and nearly half (48\%) reported frequent interaction with DeFi, NFT, or approval transactions.
Participants inspected simulated MetaMask signing screens and indicated whether they would sign or reject each request, followed by five-point ratings of perceived risk, clarity, and decision confidence. As shown in Table~\ref{tab:user_decision_summary}, sign rates ranged from 23.3\% to 70\%, revealing clear behavioral differences across task types. Requests with explicit and goal-directed purposes, such as voting (T7, 70\%) or NFT minting (T5, 60\%), generated high confidence and low perceived risk, whereas complex or ambiguous tasks such as unlimited approval (T2, 40\%) and bridging (T4, 23.3\%) yielded the lowest acceptance and highest perceived risk. Deceptive or misleading cues in suspicious contract interactions (T6, 43.3\%) and phishing-like prompts (T8, 40\%) highlighted users' reliance on heuristic trust. Prior research shows that familiar or reassuring interface language can create unwarranted assurance even in high-risk contexts~\cite{Sunshine2009CryingWolf,Akhawe2013Warningland}. Although the deceptive signatures in our study resembled legitimate prompts linguistically, participants rated them as only moderately risky, suggesting that readable but semantically uninformative messages can mask latent danger. The contrast between unlimited and regular spend requests (T2 versus T3) further showed that users respond to explicit semantic qualifiers, yet struggle to detect higher-risk variants when distinctions are encoded only in low-level parameters, a pattern consistent with difficulties in understanding permission semantics reported in prior work~\cite{Felt2012AndroidPerm,Kelley2012Conundrum}. 

Across all tasks, perceived risk showed a negative correlation with signing intention ($r=-.62$), while clarity strongly predicted decision confidence ($r=.71$). These relationships suggest that users depend more on surface recognizability and linguistic tone than on functional or permission semantics. Their mental model of what is safe to sign is shaped primarily by interface familiarity rather than by an understanding of underlying contract behavior.

\begin{table}[t]
\centering
\caption{Decision-Making Across Eight Signature Tasks}
\label{tab:user_decision_summary}
\renewcommand{\arraystretch}{1.1}
\begin{tabular}{@{}l@{\hspace{6pt}}c@{\hspace{8pt}}c@{\hspace{8pt}}c@{\hspace{8pt}}c@{}}
\toprule
\textbf{Task} & \textbf{Sign Rate (\%)} & \textbf{Risk} & \textbf{Clarity} & \textbf{Confidence} \\
\midrule
T1 – OpenSea Login & 46.7 & 3.27 & 3.27 & 4.07 \\
T2 – Unlimited Spend & 40.0 & 4.33 & 3.60 & 3.93 \\
T3 – Regular Spend & 43.3 & 2.93 & 3.53 & 3.97 \\
T4 – Bridge & 23.3 & 3.38 & 3.23 & 4.00 \\
T5 – NFT Mint & 60.0 & 2.87 & 3.63 & 3.60 \\
T6 – Suspicious Contract & 43.3 & 2.90 & 3.37 & 3.77 \\
T7 – Vote & 70.0 & 2.00 & 3.97 & 4.00 \\
T8 – Phishing Sign & 40.0 & 3.10 & 3.43 & 3.80 \\
\bottomrule
\end{tabular}
\vspace{-0.5cm}
\end{table}

\subsection{Design Goals}
\label{designgoal}

The system design was grounded in two formative investigations on how users perceive and act upon wallet signature requests. Across both studies, participants consistently struggled to connect cryptographic payloads with real-world meaning. Most attended to surface cues(token amount, recipient address, familiar layout)while overlooking functional parameters such as contract calls, chain identifiers, or expiration terms. Many described signature data as “developer code” or “unintelligible JSON,” revealing a persistent mismatch between system representation and user cognition. Even experienced users relied on heuristic trust: they favored recognizable brands or interface familiarity over contract evidence, frequently misjudging deceptive requests such as \textit{Suspicious Contract} and \textit{Phishing Sign}. Confidence often coexisted with misunderstanding, exposing a structural asymmetry in current wallet design: users can recognize what looks familiar but rarely reason about what they authorize.

Rather than redesigning the underlying signing mechanisms, we examined how semantic transparency could make existing signing interactions more interpretable and trustworthy. Three design directions guided the \textit{Signature Semantic Decoder} prototype:

\textbf{(1) Interpretive mediation.} Signing interfaces should bridge cryptographic data and human reasoning by reconstructing intent: clarifying \textit{who} is involved, \textit{what} is executed, and \textit{under what conditions}. Instead of raw parameters, users should see conceptual relations they can reason about.

\textbf{(2) Contextual risk signaling.} Because users rely on surface familiarity, risk feedback must externalize potential consequences (e.g., unlimited access, cross-contract execution) through clear visual or textual cues. Such indicators should scaffold reasoning rather than replace it.

\textbf{(3) Cognitive economy.} Interpretability must coexist with efficiency. Explanations should be concise, contextually embedded, and minimally disruptive, supporting comprehension without requiring technical expertise.

\begin{figure}[b]
\vspace{-0.2cm}
\centering
\includegraphics[width=\columnwidth]
{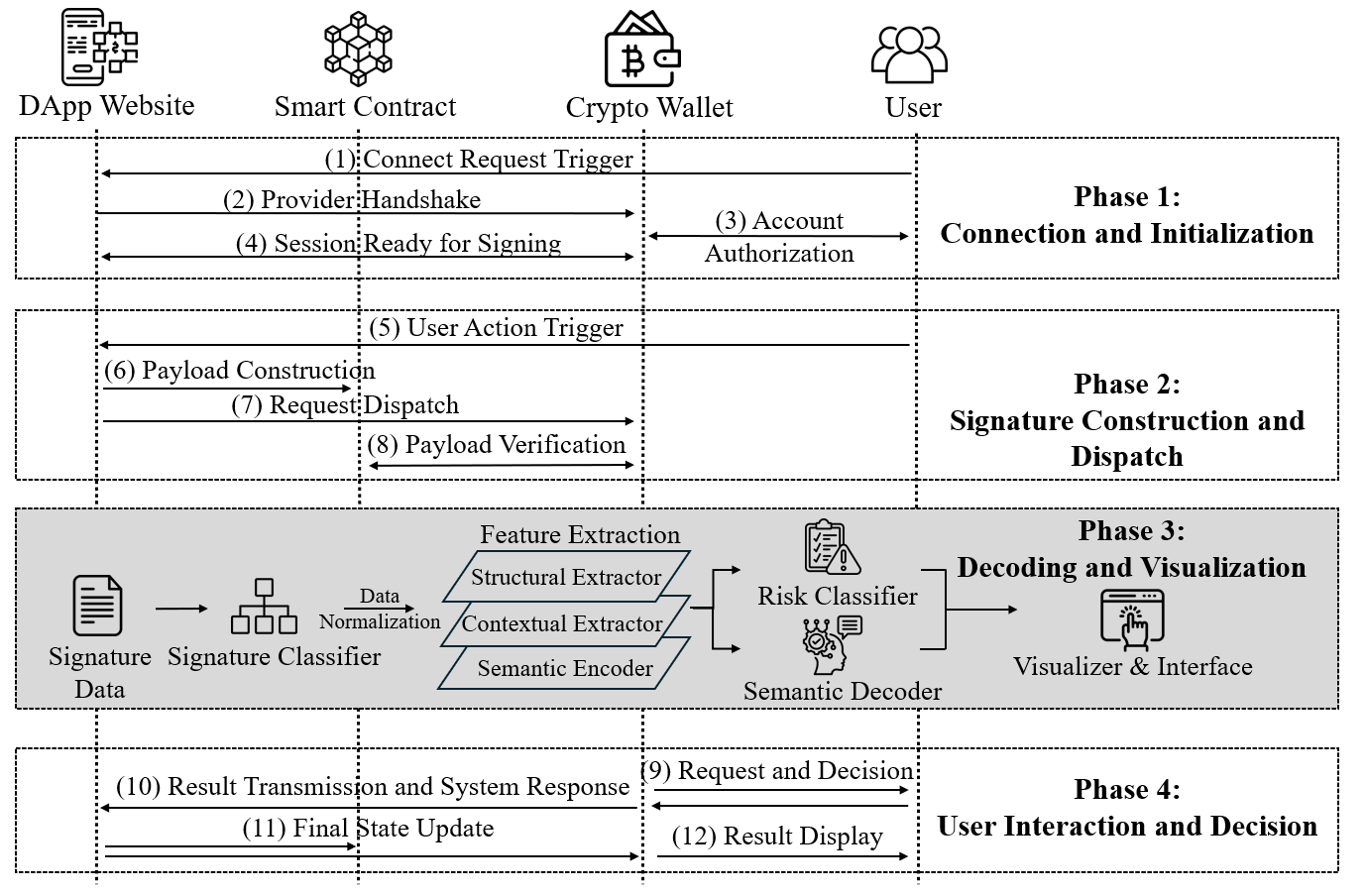}
\caption{Semantic-enhanced Signature Interaction Framework.}
\label{fig:architecture}
\end{figure}

\subsection{System Workflow}
\label{workflow}

\begin{figure}[t]
  \centering
  \vspace{-0.3cm}
  \includegraphics[width=\columnwidth]{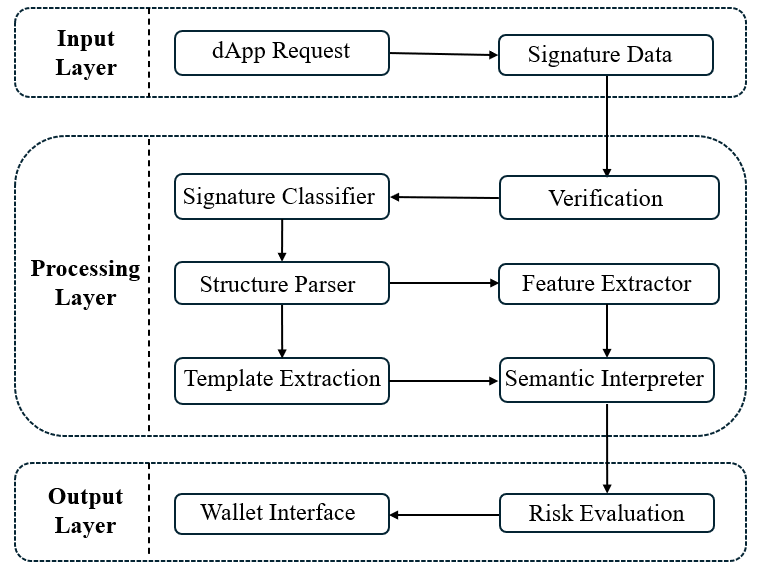}
  \vspace{-0.4cm}
  \caption{Workflow of \textit{Signature Semantic Decoder}.}
  \vspace{-0.3cm}
  \label{fig:system_workflow}
\end{figure}

The overall architecture of the system is shown in Figure~\ref{fig:architecture}. We operationalize these goals in the \textit{Signature Semantic Decoder}, a middleware layer that processes signing requests before they are rendered in the wallet interface. As illustrated in Figure~\ref{fig:system_workflow}, the architecture follows a layered pipeline comprising three main modules: the \textit{Input Layer}, the \textit{Processing Layer}, and the \textit{Output Layer}. Each layer incrementally transforms cryptographic payloads into human-readable and risk-aware narratives.

The \textit{Input Layer} receives signing triggers from decentralized applications (Dapp Request) and extracts the corresponding payloads (Signature Data). These heterogeneous inputs—\textit{eth\_sendTransaction}, \textit{personal\_sign}, or \textit{eth\_signTypedData}—carry distinct syntactic and contextual structures. To ensure consistent interpretation, the system performs schema validation and domain separation before parsing, filtering malformed or replayed payloads. Structured parsing then normalizes the input: transaction-type requests are decomposed into sender, recipient, and value fields, while typed-data requests are recursively expanded according to their EIP-712 schemas. The result is a unified intermediate representation that abstracts low-level formatting differences while preserving semantic attributes for downstream reasoning.

Within the \textit{Processing Layer}, the \textit{Signature Classifier} determines category and verifies integrity, while the \textit{Structure Parser} reconstructs relations—identifying who acts, on what, and under what conditions. Extracted features such as token type, contract address, and function selector are mapped to human-understandable roles (“spender,” “token contract,” “approval limit”) via a hybrid method combining rule-based templates and ABI-assisted inference. The \textit{Semantic Interpreter} aggregates these mappings to infer high-level intent, scope, and delegation effects, linking them to a knowledge base of contract patterns to flag latent risks like unlimited approvals or hidden delegations. This process converts machine-level parameters into a structured semantic frame verbalized for end users.

The \textit{Output Layer} converts this frame into interpretable feedback. The \textit{Risk Evaluation} module quantifies exposure based on action scope, contract reputation, and context, while the \textit{Visualization and Feedback} component renders plain-language summaries and color-coded indicators within the wallet interface. Natural-language templates generate concise explanations—e.g., “You are allowing Contract X to spend up to 100~USDC on your behalf until expiration.” These textual cues, paired with visual signals (icons), communicate risk levels and underlying rationale, turning cryptographic data into coherent narratives that support informed, trustworthy signing decisions.

While the prototype implements the complete interpretive pipeline shown in Figure~\ref{fig:system_workflow}, its purpose is illustrative rather than exhaustive. The system was developed as a research prototype to demonstrate the feasibility of semantic decoding and to support user evaluation, rather than to optimize runtime performance or protocol coverage. Technical implementation details such as ABI parsing, schema validation, and natural-language processing are abstracted here, as the study focuses on interpretability and decision support rather than engineering performance. Further technical implementation details are available in the accompanying open-source prototype\footnote{The code will be made publicly available upon publication.}.

\begin{figure*}[h]
\vspace*{-1cm}
\centering
\includegraphics[width=\textwidth]{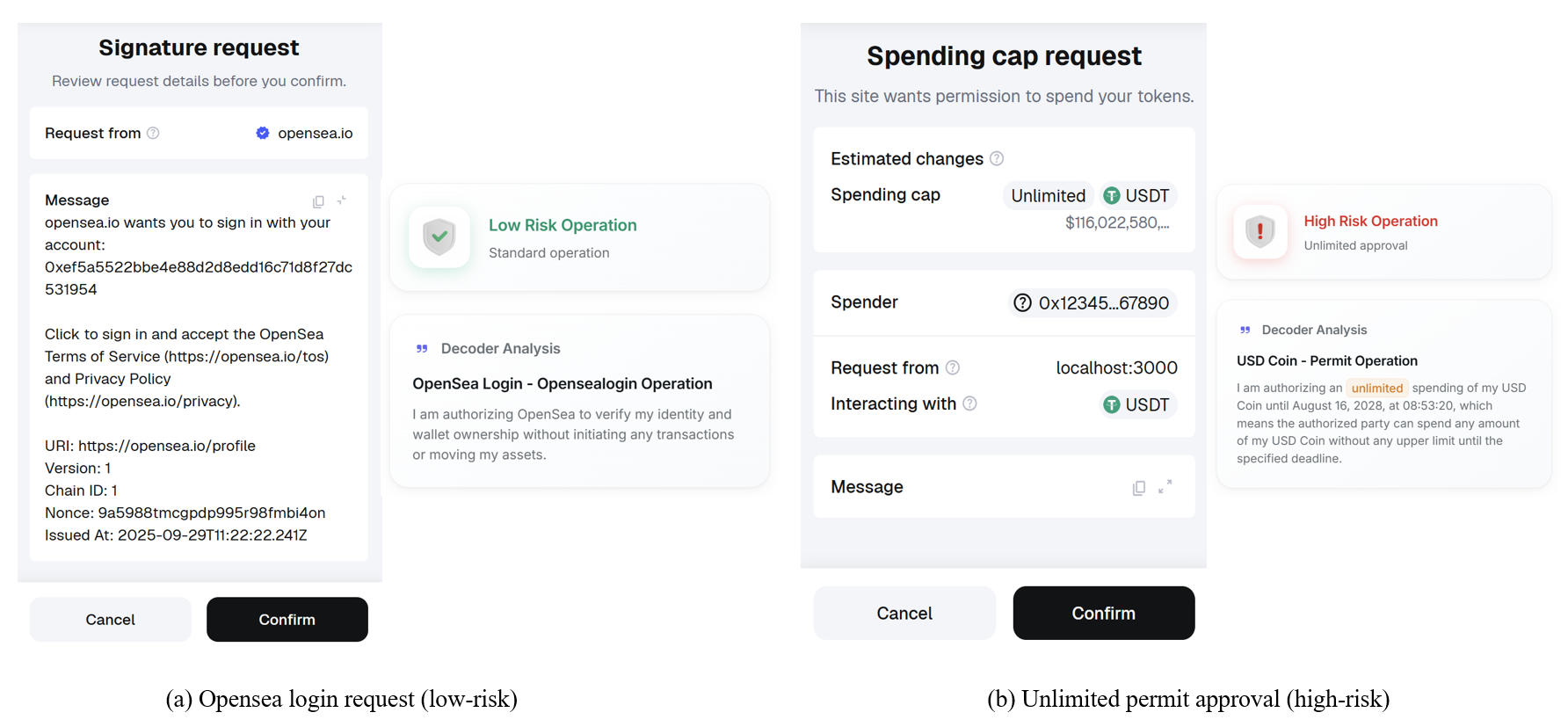}
\caption{MetaMask-based interface examples from the Signature Semantic Decoder}
\label{fig:interface}
\end{figure*}

\begin{figure*}[h]
\vspace{-0.3cm}
\centering
\includegraphics[width=\textwidth]{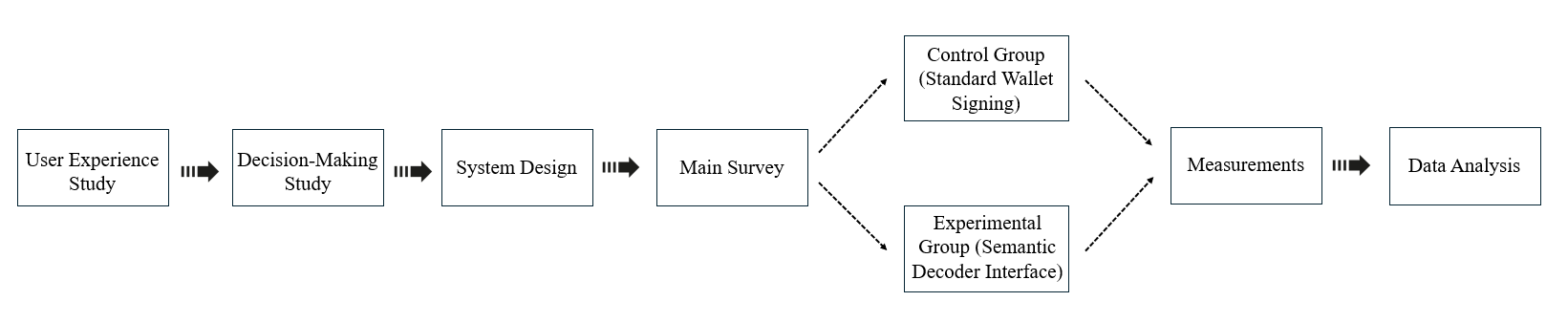}
\caption{Overall Research Pipeline.}
\label{fig:research_workflow}
\vspace{-0.3cm}
\end{figure*}

\subsection{User Interface and Visualization}
\label{ui}

The user interface constitutes the final stage of the \textit{Signature Semantic Decoder}, where reconstructed intents and inferred risks are presented in a human-readable, visually interpretable form. Building upon the preceding semantic modules, this layer translates abstract cryptographic payloads into clear, contextual explanations that directly support signing decisions. The design objective is to make semantic transparency visible at the point of interaction, enabling users to understand what they authorize without disrupting their normal workflow.

The interface follows a top–down information hierarchy derived from attention patterns observed in the formative studies. At the top, a plain-language summary concisely conveys the primary intent of the signature request—for example, “You are granting Uniswap permission to spend up to 100 USDC on your behalf.” This statement answers the question users most often ask: \textit{“What will happen if I sign?”} Beneath it, structured details such as token type, spender address, and deadline are displayed in a static contextual panel.Important fields that may influence security decisions are additionally highlighted to draw immediate attention.

Risk information inferred by the reasoning engine is displayed alongside the semantic summary through a three-tier color scheme—green for low, yellow for moderate, and red for high-risk operations—accompanied by icons and brief tooltips that clarify the rationale (e.g., “Unlimited approval detected: spender may access your entire token balance”). This visual composition promotes calibrated trust by revealing the logic behind each warning rather than issuing generic alerts. The interface adopts a minimalist aesthetic with balanced typography, spacing, and iconography to maintain clarity under cognitive load, while accessibility features such as color contrast and shape redundancy follow established usability and readability guidelines~\cite{Nielsen1994Usability,WCAG21,Sweller1988CLT,Wong2011Color,BravoLillo2011}.

As shown in Figure~\ref{fig:interface}, the interface extends the native MetaMask signing view with an interpretive side panel. Each request preserves the standard wallet layout for compatibility while embedding decoded explanations and risk cues. In the OpenSea login example (Figure 5a), the system recognizes a routine authentication and labels it as a low-risk operation with a green badge and brief rationale. In contrast, the USDC permit approval (Figure 5b) exposes an unlimited token allowance and is marked as high risk with red cues and explanatory text describing its persistent authorization. These examples illustrate how the \textit{Signature Semantic Decoder} augments existing wallet workflows, turning opaque signing prompts into semantically interpretable and trustworthy decisions.

\section{User Study}

This section presents the user study designed to evaluate the effectiveness of the \textit{Signature Semantic Decoder} in improving users’ comprehension, confidence, and trust when responding to crypto wallet signature requests.

\subsection{Study Design}
\label{sec:study_design}

As illustrated in Figure~\ref{fig:research_workflow}, the user study served as the summative evaluation phase of our research pipeline. It aimed to assess whether the \textit{Signature Semantic Decoder} (SSD) could enhance users’ comprehension, confidence, and trust when responding to wallet signature requests. Insights from the two preceding formative studies informed the construction of realistic signing scenarios, the definition of interpretive cues, and the choice of dependent measures. The experiment adopted a between-subjects design~\cite{Lazar2017HCIResearch} comparing two wallet interfaces: a baseline replicating MetaMask’s default signing screen and a semantic-enhanced version integrating our research prototype.

\textbf{Implementation and Conditions.}
We developed a browser-based wallet simulator that replicated the structure and flow of MetaMask’s confirmation interface while running entirely offline to eliminate privacy or financial risk. All signing payloads were pre-generated from authentic Ethereum contracts covering common DeFi and NFT use cases (e.g., token approval, cross-chain bridge, governance vote). The simulator supported three major signing formats—\textit{eth\_sendTransaction}, \textit{personal\_sign}, and \textit{eth\_signTypedData} (EIP-712)~\cite{EIP712}—allowing direct comparison across on-chain and off-chain paradigms.  
In the semantic-enhanced condition, each payload was parsed by the SSD to reconstruct actor–action–object relations, infer intent, and generate natural-language explanations accompanied by color-coded risk cues derived from the model’s internal reasoning module. Both conditions maintained identical layouts, fonts, button placements, and color palettes to isolate interpretability effects from visual or interactional differences. Interaction latency and page layout were pretested to ensure consistent usability across conditions.

\textbf{Task Corpus and Risk Classification.}
Each participant completed six signing tasks representing common wallet operations observed across decentralized applications. The task set covered: (T1) \textit{Opensea Login} via \textit{personal\_sign}, (T2) \textit{NFT Mint} through a \textit{transaction} call, (T3) \textit{DAO Vote} using EIP-712 structured data, (T4) \textit{Bridge/Swap} for cross-contract token transfer, (T5) \textit{Unlimited Approval} granting full-spend permissions, and (T6) a simulated \textit{Phishing Request} mimicking a malicious signature prompt. These scenarios were drawn from a curated corpus of authentic Ethereum transaction payloads collected from open-source dApps~\cite{DappRadar}, stripped of branding information, and re-rendered into a standardized template to maintain visual consistency.  
Each task contained a single signature request differing in semantic structure, permission scope, and potential asset impact. To probe users’ sensitivity to varying risks, tasks were pre-classified as low, medium, or high risk based on their underlying contract logic and privilege escalation potential. Specifically, the \textit{Opensea Login} and \textit{NFT Mint} scenarios were labeled as low-risk, the \textit{DAO Vote} and \textit{Bridge/Swap} tasks as medium-risk, and the \textit{Unlimited Approval} and \textit{Phishing Request} tasks as high-risk. Classification was determined through expert review of transaction parameters, function signatures, and authorization scope, achieving substantial inter-rater agreement ($\kappa = 0.82$). This taxonomy operationalized “risk” as a latent permission gradient rather than a visual cue, enabling precise comparison between objective task structure and subjective risk perception.

\textbf{Interaction Flow.}
Participants were randomly assigned to one of the two interface conditions and completed six tasks presented in randomized order~\cite{Lazar2017HCIResearch} to minimize sequence and fatigue effects. Each task displayed a simulated signing interface: the baseline condition replicated MetaMask’s native confirmation view, presenting the raw hexadecimal or structured key–value data that wallets typically reveal, while the semantic-enhanced version applied our semantic decoder to reconstruct actor–action–object relations, generate a plain-language intent summary, highlight critical fields, and provide contextual explanations together with color-coded risk cues adjacent to the \textit{Sign} button. After each decision to \textit{sign} or \textit{reject}, participants rated perceived risk, clarity, and confidence on five-point Likert scales~\cite{Joshi2015Likert}. These behavioral and perceptual data were logged automatically to compute decision accuracy against the pre-assigned risk classification, providing both objective and subjective indicators of comprehension and caution.  

\textbf{Procedure.} The study was conducted online via a secure survey platform and lasted approximately 15–20 minutes. After providing informed consent, participants were informed about the study objectives, data-handling policies, and their right to withdraw at any time. A short tutorial introduced the simulated wallet interface, explained the concept of signature requests, and demonstrated how to make and record signing decisions. Participants then completed one practice task to become familiar with the interaction flow and the post-decision rating procedure. During the main session, each participant completed six randomized signing scenarios that represented diverse Web3 actions such as approvals, delegations, and mints. Each scenario began with a simulated wallet confirmation screen showing transaction details; participants could explore and review all on-screen information before deciding whether to \textit{sign} or \textit{reject}. After each decision, they rated perceived risk, clarity, and confidence using five-point Likert scales. All responses, timestamps, and decision outcomes were automatically logged to enable later computation of accuracy, deliberation time, and behavioral consistency. To replicate the natural rhythm of wallet interactions, no time limits were imposed, but a soft progress indicator was provided to maintain engagement. Upon completing all six tasks, participants answered a brief demographic questionnaire and provided short open-ended reflections describing how they assessed risks, what cues influenced their confidence, and how the interface compared to their prior wallet experience.

\subsection{Recruitment and Sample}
\label{sec:recruitment_sample}

Participants were recruited via Prolific without geographic restrictions, limited to individuals who had previously held cryptocurrency and used a non-custodial wallet (e.g., MetaMask, Phantom, or Trust Wallet) to connect to decentralized applications. Screening questions verified wallet familiarity and excluded respondents who had never signed a blockchain transaction. Submissions with duplicate IP addresses, failed attention checks, or implausible completion times (beyond 1.5×IQR, where IQR denotes the interquartile range) were removed, ensuring only attentive and experienced users were retained. An a priori power analysis using G*Power~\cite{Faul2007GPower} estimated a minimum sample of $N=128$ (64 per condition) to detect medium effects ($d=0.5$, $\alpha=0.05$, 80\% power). After applying attention checks, completion time filters, and duplicate screening, 128 valid responses (64 per condition) remained, yielding statistical power above 0.80.
Participants were randomly assigned to either the baseline or semantic-enhanced interface and completed the online study at their own pace, with a median duration of 16 minutes and compensation rates around £9/hour, meeting Prolific’s fair-pay standard. Participant demographics are presented in Table~\ref{tab:appendix_demographics} in the Appendix~\ref{Appendix:participant_table}. Participation was voluntary and anonymous, with no real wallet connections, on-chain transactions, or personal identifiers collected.

\subsection{Measures and Data Analysis}
\label{sec:measures_analysis}

Participants’ performance was evaluated using both behavioral measures and subjective self-reports to examine how interface type and risk level influenced comprehension, confidence, and workload. Objective performance was based on participants’ \textit{sign} or \textit{reject} decisions across six tasks, with decision accuracy defined as the proportion of responses consistent with each task’s ground-truth risk classification~\cite{Sunshine2009CryingWolf,Akhawe2013Warningland}—correctly rejecting high-risk signatures and approving low-risk ones. Task completion time was recorded as an exploratory measure of decisional effort. Subjective measures included per-task ratings of perceived \textit{risk}, \textit{clarity}, and \textit{confidence} on five-point Likert scales, as well as post-study scales assessing cognitive workload (NASA–TLX~\cite{Hart1988TLX}), trust in automation~\cite{Jian2000Trust}, and perceived comprehensibility, each adapted from validated instruments~\cite{Kulesza2013Principles} and showing high internal consistency (Cronbach’s~$\alpha > .80$). Quantitative data were screened for completeness and normality, and analyzed using two-way mixed ANOVAs with \textit{interface type} (baseline vs.~semantic-enhanced) as a between-subject factor and \textit{risk level} (low, medium, high) as a within-subject factor. Dependent variables included decision accuracy, perceived risk, clarity, and confidence; post-study scales were compared with independent-samples $t$-tests, and effect sizes reported as Cohen’s~$d$ or partial~$\eta^2$~\cite{Cohen1988EffectSize}. Correlation analyses explored relationships among trust, comprehensibility, and workload to examine potential trade-offs between interpretability and cognitive effort. Open-ended reflections were thematically coded by two independent researchers, with intercoder reliability verified using Cohen’s~$\kappa$~\cite{Landis1977Kappa}, providing complementary qualitative insight into how semantic decoding shaped users’ reasoning and trust during signing.

\subsection{Limitations}
While the controlled, simulation-based setup ensured high internal validity, it inevitably differed from real-world signing contexts where users face genuine financial stakes, time pressure, and emotional engagement. This abstraction enabled precise isolation of cognitive effects but may limit ecological validity, as participants interacted with simulated wallet screens rather than executing live transactions. Consequently, behavioral nuances such as hesitation, verification habits, or post-decision regret common in high-value scenarios may not have been fully captured. The between-subjects design offered a clear comparison between interfaces and prevented carry-over effects but reduced statistical power relative to within-subject approaches, limiting analysis of individual learning and adaptation. Although the power analysis confirmed sufficient sensitivity for medium effects, smaller or interaction-level effects might remain undetected. Future studies could employ within-subject or longitudinal designs with real or incentivized assets to examine how interpretability evolves with repeated exposure and under genuine financial conditions. Moreover, while the Prolific recruitment strategy yielded a heterogeneous pool across regions and experience levels, participants were predominantly English-speaking retail users. Professional traders or users operating under different regulatory, linguistic, or cultural contexts may perceive signing risks differently. Extending evaluations to mobile and in-situ environments, where multitasking and contextual distractions occur, would further test how semantic transparency performs under authentic operational constraints and behavioral complexity.

\subsection{Ethical Considerations}
The study protocol was reviewed and approved by the university’s ethics committee and conducted in accordance with the Menlo Report~\cite{Menlo2012} and relevant data protection regulations~\cite{GDPR2016}. Participants provided informed consent before participation and were clearly informed that all interactions involved simulated signing screens only. No real blockchain transactions, private-key operations, or identifiable wallet data were used, ensuring that participants faced no financial or privacy risks. Demographic questions were optional, and all responses were anonymized, encrypted, and stored on secure institutional servers accessible only to the research team. Participation was voluntary, and participants could withdraw at any time without penalty. Compensation followed Prolific’s fair-pay policy~\cite{ProlificFairPay} (approximately £9/hour), ensuring equitable remuneration without undue inducement. No adverse events, complaints, or data-related issues were reported during the study, and only aggregated, anonymized results were used for analysis and replication.

\newcommand{\riskL}{\textcolor{green!50!black}{L}} % Low risk
\newcommand{\riskM}{\textcolor{orange!90!black}{M}} % Medium risk
\newcommand{\riskH}{\textcolor{red!70!black}{H}} % High risk

\begin{table*}[t]
\vspace{-1cm}
\centering
\caption{Participant Decisions and Perception Ratings Across Six Tasks}
\label{tab:task_results}
\renewcommand{\arraystretch}{1.1}
\setlength{\tabcolsep}{4.7pt}
\footnotesize
\begin{tabularx}{\textwidth}{@{}l|
  >{\centering\arraybackslash}p{1cm}|
  >{\centering\arraybackslash}p{1cm}|
  cccc|cccc@{}}
\toprule
 &  &  & \multicolumn{4}{c|}{\textbf{Control Group (N=64)}} & \multicolumn{4}{c}{\textbf{Experimental Group (N=64)}} \\
\textbf{Task} & \textbf{Method} & \textbf{Risk} & \textbf{Sign\%} & \textbf{Risk} & \textbf{Clarity} & \textbf{Conf.} & \textbf{Sign\%} & \textbf{Risk} & \textbf{Clarity} & \textbf{Conf.} \\
\midrule
T1 — Opensea Login & PS & \riskL &
61.5 & 2.87 (1.46) & 3.42 (1.45) & 3.88 (1.18) &
86.3 & 1.88 (1.23) & 4.16 (1.10) & 4.27 (0.90) \\
T2 — NFT Mint & TX & \riskL &
57.7 & 3.23 (1.23) & 3.38 (1.29) & 3.88 (1.02) &
92.2 & 1.98 (1.09) & 4.16 (0.90) & 4.25 (0.72) \\
T3 — DAO Vote & E712 & \riskM &
69.2 & 2.71 (1.18) & 3.29 (1.19) & 3.83 (1.17) &
66.7 & 2.94 (0.88) & 3.61 (1.18) & 3.84 (0.95) \\
T4 — Bridge / Swap & E712 & \riskM &
73.1 & 2.71 (1.27) & 3.69 (1.29) & 3.83 (1.20) &
51.0 & 3.14 (1.02) & 3.80 (0.94) & 4.02 (0.81) \\
T5 — Unlimited Approval & E712 & \riskH &
36.5 & 3.63 (1.40) & 3.38 (1.33) & 4.00 (1.03) &
9.8 & 4.73 (0.70) & 4.00 (1.22) & 4.31 (1.05) \\
T6 — Phishing Request & E712 & \riskH &
32.7 & 3.54 (1.21) & 3.31 (1.39) & 3.90 (1.26) &
17.7 & 4.39 (1.04) & 3.61 (1.28) & 4.37 (0.92) \\
\bottomrule
\end{tabularx}
\parbox{\textwidth}{
\vspace{1mm}
\scriptsize
\textit{In this table,} PS = \texttt{personal\_sign}, 
E712 = \texttt{eth\_signTypedData\_v4} (EIP-712 typed data), 
TX = \texttt{eth\_sendTransaction}.

\textit{Risk levels:} 
\textcolor{green!50!black}{L} = Low; 
\textcolor{orange!90!black}{M} = Medium; 
\textcolor{red!70!black}{H} = High;
Values in parentheses indicate standard deviations.
}
\end{table*}

\begin{figure*}[t]
\centering
\includegraphics[width=\textwidth]{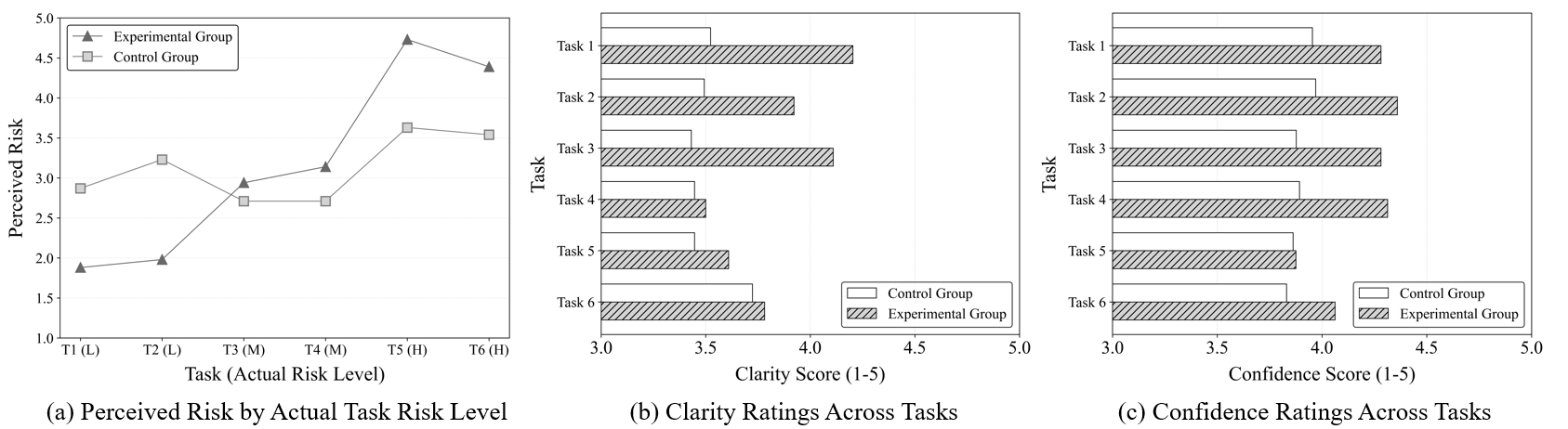}
\caption{Comparison of participants’ perception and comprehension measures between two groups.}
\label{fig:comparison}
\end{figure*}

\section{Results}
The main study compared a baseline wallet interface with an enhanced version incorporating semantic decoding and visualization across six simulated signing tasks of varying risk levels.
Of the 128 participants who completed the study, 64 were assigned to the control group and 64 to the experimental group (Appendix~\ref{Appendix:participant_table}.Table~\ref{tab:appendix_demographics}). 
Most were experienced crypto users, with the majority aged between 25 and 34 years and 66\% reporting an IT-related background. 
No significant demographic differences were found between groups, indicating comparable samples. In the following, we will present our findings to the research questions.

\subsection{Decision Accuracy and Task Performance}

\begin{tcolorbox}[colback=gray!5,colframe=black!30,arc=2mm,boxrule=0.4pt]
RQ1: To what extent does semantic transparency improve users’ ability to identify deceptive or risky signatures without increasing cognitive workload?
\end{tcolorbox}

To address \textbf{RQ1}, we examined how semantic decoding and visualization influenced users’ ability to correctly identify risky or deceptive signing requests relative to a baseline wallet. Analyses included objective decision accuracy, perceived risk, and task efficiency across six representative signing scenarios spanning low-, medium-, and high-risk levels (Table~\ref{tab:task_results}).

Participants using the semantic interface achieved substantially higher accuracy in distinguishing safe from unsafe transactions. Across all tasks, the experimental group reached an average correct identification rate of 84.2\% ($SD$=8.7) compared to 67.9\% ($SD$=11.5) in the control group, a large and significant difference ($t$(102)=6.41, $p<.001$, $d$=0.86, 95\% CI [0.58, 1.13]). A two-way mixed ANOVA with interface type and risk level revealed a main effect of interface ($F$(1,100)=40.9, $p<.001$, $\eta^2_p$=.29) and a significant interaction ($F$(2,200)=8.31, $p<.001$)~\cite{Field2013Statistics}, indicating that the semantic interface’s advantage increased with task risk. Improvements were particularly pronounced in high-risk scenarios: false approval rates for \textit{Unlimited Approval} (T2) and \textit{Delegate/Vote} (T4) dropped from 36.5\% and 32.7\% to 9.8\% and 17.7\%, respectively ($\chi^2$(1, $N$=103)=9.71, $p<.01$). For low- and medium-risk transactions (T1, T3, T5, T6), both groups maintained accuracy above 85\%, confirming that interpretive transparency improved discrimination without inflating caution.

As shown in Figure~\ref{fig:comparison}(a), perceived risk ratings in the experimental condition closely followed the objective task hierarchy, increasing linearly from 1.93 (low) to 4.42 (high), while control-group ratings remained comparatively flat (3.05–3.31). This difference was confirmed by a repeated-measures ANOVA ($F$(2,100)=27.8, $p<.001$, $\eta^2_p$=.36)~\cite{Field2013Statistics}, revealing a sharper gradient of perceived risk across levels. Bayesian analysis further supported this calibration effect ($BF_{10}>100$)~\cite{Rouder2009Bayes,Wagenmakers2017JASP}, suggesting that semantic risk indicators effectively aligned subjective judgments with true task severity, consistent with risk-perception theory~\cite{Slovic1987Perception,Tversky1981Framing}. Subjective comprehension and confidence also improved: clarity ratings rose from $M$=3.35 ($SD$=1.33$)$ to $M$=3.91 ($SD$=1.03$)$ ($t$(102)=2.18, $p<.05$, $d$=0.45), and confidence from $M$=3.85 ($SD$=1.08$)$ to $M$=4.12 ($SD$=0.94$;$ $p<.05$). Participants frequently noted that “the plain-language summary made the purpose obvious” and that “the warning icon prevented impulsive confirmation,” indicating heightened self-awareness and deliberation during signing.

\begin{table}[t]
\centering
\caption{Task Completion Time Comparison Between Groups}
\label{tab:task_completion_time}
\renewcommand{\arraystretch}{1.05}
\setlength{\tabcolsep}{5pt}
\begin{tabular}{l|S|S|S|S}
\toprule
\textbf{Task} & \textbf{Control (s)} & \textbf{Experimental (s)} & \textbf{Diff. (s)} & \textbf{Change (\%)} \\
\midrule
Task 1 & 45.02 & 50.11 & -5.09 & +11.31 \\
Task 2 & 54.01 & 49.52 & +4.49 & -8.31 \\
Task 3 & 58.16 & 51.60 & +6.55 & -11.27 \\
Task 4 & 50.52 & 59.65 & -9.13 & +18.08 \\
Task 5 & 54.01 & 53.26 & +0.75 & -1.39 \\
Task 6 & 53.38 & 56.55 & -3.17 & +5.94 \\
\midrule
\addlinespace[2pt]
\textbf{Mean} & \textbf{52.52} & \textbf{53.45} & \textbf{-0.93} & \textbf{+1.78} \\
\bottomrule
\end{tabular}
\vspace{-0.5cm}
\end{table}

Completion times (Table~\ref{tab:task_completion_time}) were modestly longer in the semantic condition ($M$=53.6s vs.\ 46.8s; $t$(102)=2.02, $p=.046$, $d$=0.39), mainly in high-risk tasks where participants paused to read explanations. However, within-condition variance decreased ($F$-ratio=1.87, $p<.05$), suggesting more consistent decision pacing. Correlation analysis showed that perceived clarity positively predicted accuracy ($r$=.61, $p<.001$) and negatively predicted completion time ($r$=-.42, $p=.002$), implying that added deliberation reflected focused reasoning rather than confusion~\cite{Sweller1988CLT,Kulesza2013Principles}. Control participants frequently approved deceptive signatures due to superficial familiarity with addresses or tokens, whereas experimental participants’ errors were conservative—typically classifying medium-risk tasks as high-risk or hesitating before signing. This shift from careless acceptance to cautious evaluation aligns with prior findings on habituation and superficial processing in security decisions~\cite{Sunshine2009CryingWolf,Akhawe2013Warningland} and suggests a recalibration of risk discrimination.

The benefits of semantic decoding were concentrated in deceptive high-risk tasks (T5, T6), yielding relative risk reductions of 73\% and 46\%. For \textit{personal\_sign} and simple transfer tasks, effects were moderate or neutral, demonstrating generalizability across signing paradigms without usability degradation. A post-hoc Tukey test confirmed significant differences between high- and low-risk categories ($p<.001$) but not between low- and medium-risk ones, indicating selective enhancement of vigilance rather than uniform caution.

Rather than making users trust the system by default, the semantic interface enabled them to \textit{justify} their trust. Participants’ confidence was grounded in understanding rather than delegation—they signed because they knew what they were authorizing, not because the system claimed safety. This reframing of trust from automation to interpretability is consistent with foundational research on calibrated trust in automation and explainable systems~\cite{Lee2004Trust,Abdul2018XAI}. These observations motivate the following section, which examines how such interpretive transparency shapes users’ perceptions of trust and system reliability in practice.

\subsection{Perceived Understanding and Trust}
\begin{tcolorbox}[colback=gray!5,colframe=black!30,arc=2mm,boxrule=0.4pt]
RQ2: How does semantic transparency affect users’ comprehension, confidence, and perceived control when evaluating wallet signature requests?
\end{tcolorbox}

\begin{figure}[t]
    \vspace{-0.4cm} % 减少与上方文字的距离
    \centering
    \includegraphics[width=\columnwidth]{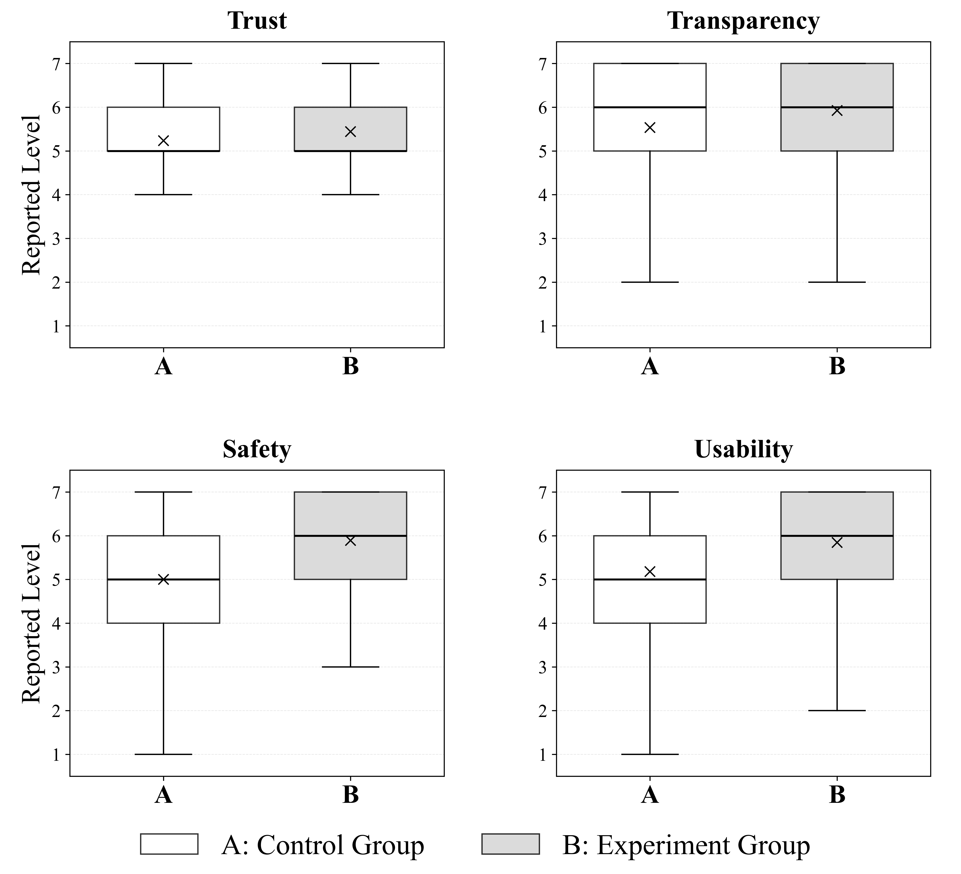}
    \vspace{-0.5cm}
    \caption{Post-study ratings of Trust, Transparency, Safety, and Usability.}
    \label{fig:ratings}
    \vspace{-0.3cm}
\end{figure}

\begin{figure}[t]
    \centering
    \includegraphics[width=\columnwidth]{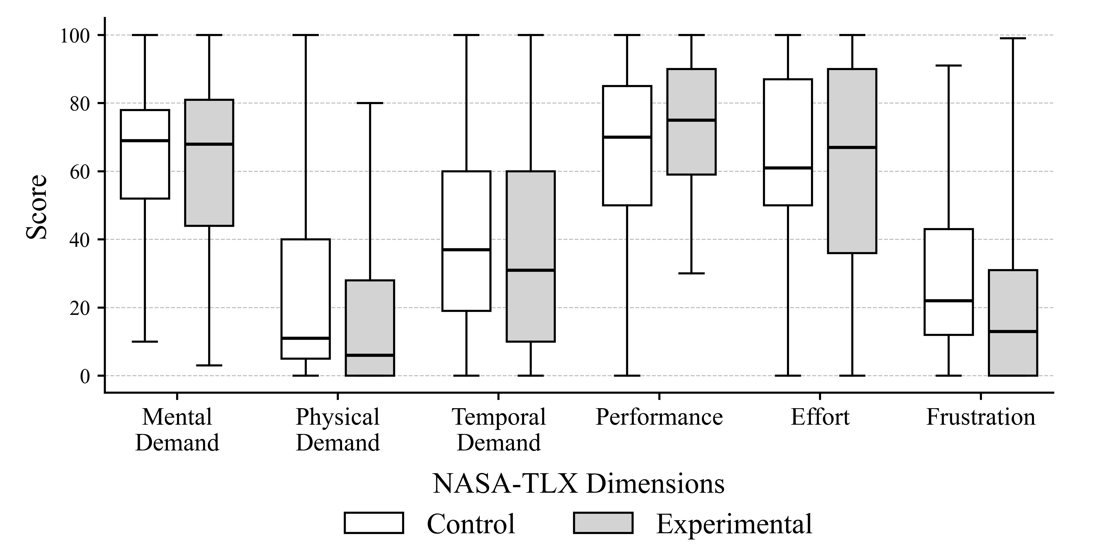}
    \vspace{-0.5cm}
    \caption{NASA-TLX Ratings.}
    \label{fig:nasa_tlx_boxplot}
    \vspace{-0.5cm}
\end{figure}

To address \textbf{RQ2}, we examined how semantic decoding and visualization affected participants’ perceived comprehension, trust, and workload. Post-study ratings of clarity, transparency, safety, and usability (Figure~\ref{fig:ratings}) were analyzed alongside NASA–TLX workload scores (Figure~\ref{fig:nasa_tlx_boxplot}).

Participants using the semantic interface reported significantly greater comprehension of signing intent and clearer perception of transaction parameters. Mean clarity ratings (“the meaning of the signature request was clear to me”) increased from $M$=5.46 ($SD$=1.32) to $M$=6.13 ($SD$=0.98) ($t$(127)=3.11, $p<.01$, $d$=0.55), and understanding of post-signing consequences rose from $M$=5.69 to $M$=6.32 ($t$(127)=2.96, $p<.01$). Participants also agreed more strongly that key elements such as amount, allowance, chain, and deadline were clear ($M$=6.11$>$5.48; $p<.01$) and that they could easily identify potential risks ($M$=5.91$>$5.13; $p<.001$). Qualitative remarks reinforced these results: several participants noted that “the summaries told me what the transaction really does” and that “risk icons showed why one was dangerous even if the text looked similar.”

\begin{table*}[t]
\vspace{-1cm}
\centering
\caption{Thematic Analysis of Experimental Group}
\label{tab:semantic_helpfulness}
\renewcommand{\arraystretch}{1.15}
\setlength{\tabcolsep}{5pt}
\footnotesize
\begin{tabularx}{\textwidth}{|p{2.2cm}|>{\centering\arraybackslash}X|X|X|}
\hline
\multicolumn{1}{|l|}{\textbf{ID Range}} &
\multicolumn{1}{c|}{\textbf{Theme}} &
\multicolumn{1}{c|}{\textbf{Definition}} &
\multicolumn{1}{c|}{\textbf{Quote}} \\
\hline

1--32, 35, 38, 42--44, 47--48, 50--51, 53--57, 61, 64 
& Risk Cues and Level Indicators 
& Risk labels and color cues enabled quick assessment of request safety. 
& ``The warning that states if it's high risk or medium risk was very helpful.'' \\
\hline

4--5, 10--12, 15, 19, 36, 39, 52, 55, 58 
& Semantic Interpretation 
& Summaries clarified intent and explained the effect of each signature. 
& ``It explained in simple words what would happen after signing.'' \\
\hline

8--9, 14, 18, 20, 34, 45, 60 
& Layout and Comprehensibility 
& The interface was organized but visually dense, requiring focused reading. 
& ``It was clear overall but looked crowded and busy.'' \\
\hline

1--2, 10, 15, 21, 23 
& Transparency and Context Visibility 
& Showing addresses, tokens, and context increased clarity and confidence. 
& ``I liked how everything was shown, including both sender and receiver wallets.'' \\
\hline

7, 13, 14, 17, 23, 61, 64 
& Enhanced Security Awareness and Trust 
& Risk cues strengthened users’ sense of control and perceived safety. 
& ``It felt safer because I knew the risk level.'' \\
\hline

7, 9, 11, 36, 59--63 
& Lack of Detailed Explanation 
& Users wanted clearer reasons behind medium or low risk ratings. 
& ``If it was medium or low, I didn’t know the reason or what made it risky.'' \\
\hline

\end{tabularx}
\end{table*}

Improved comprehension translated directly into higher trust and perceived reliability~\cite{Kulesza2013Principles,Hoffman2018Explanation}. Experimental-group participants reported greater confidence in the system’s interpretation ($M$=5.98$>$5.23; $p=.007$), reliability ($M$=5.89$>$5.34; $p=.009$), and willingness to rely on it in future signing ($M$=5.93$>$5.18; $p=.004$), and more strongly agreed that the interface helped them avoid harmful actions ($M$=6.28$>$5.43; $p<.001$). These findings indicate that semantic interpretability cultivates \textit{calibrated trust}: users trusted the system because they could verify its reasoning, not because it automated decisions~\cite{Lee2004Trust,Abdul2018XAI}.

Figure~\ref{fig:ratings} shows that trust, transparency, safety, and usability ratings were consistently higher and less variable in the semantic condition, with median scores of 6 (out of 7) for transparency and safety versus 5 in the control group. Participants noted that the explanations “removed the guesswork’’ and that “seeing structured data explained in plain English made it feel safer to sign,’’ with several adding that the interface helped them verify intent rather than rely on intuition alone.

NASA–TLX analysis further revealed lower cognitive workload without added effort. Mental demand decreased from $M$=67.8 to $M$=42.6 ($t$(127)=4.12, $p<.001$, $d$=0.73), and frustration from $M$=61.2 to $M$=38.7 ($t$(127)=3.97, $p<.001$), while physical and temporal demands remained low. Participants described the slight increase in task time as intentional checking rather than confusion, reporting that the explanations made their decisions “feel more deliberate.’’ Semantic decoding thus mitigated cognitive overload by translating cryptographic data into actionable, human-readable information~\cite{Sweller1988CLT}.

Interpretability shapes both the cognitive and emotional dimensions of the signing process. By transforming abstract cryptographic data into coherent, human-readable narratives, the semantic interface enabled participants to approach transactions with greater understanding and control~\cite{Hoffman2018Explanation}. Trust stemmed from comprehension rather than automation, as participants gained confidence through awareness of what they were authorizing while maintaining agency~\cite{Lee2004Trust}. This balance of clarity, confidence, and reduced cognitive demand underscores explanation as a foundation of trustworthy human–system interaction~\cite{Abdul2018XAI}.

\subsection{Evaluation of System Effectiveness}
\begin{tcolorbox}
[colback=gray!5,colframe=black!30,arc=2mm,boxrule=0.4pt]
RQ3: How do users perceive the role of semantic transparency in shaping their trust and overall signing experience?
\end{tcolorbox}

Participants’ reflections (Table~\ref{tab:semantic_helpfulness}) illustrate how the Signature Semantic Decoder reshaped their sense of comprehension, control, and emotional assurance during the signing process. Beyond the quantitative gains in accuracy and confidence, the qualitative responses offer a grounded account of what participants noticed, relied on, or struggled with when interpreting signature requests at scale.

Risk cues were the most frequently mentioned element, appearing across users with diverse experience levels. Participants described the color-coded warnings as “the first thing I look at” (P7), “a quick way to know if I should be careful” (P22), and “the part that tells me how seriously to read the rest” (P31). Several noted that the labels (“high,” “medium,” “low”) provided an immediate sense of structure: “The warning that states if it’s high risk or medium risk was very helpful” (P3), and “Seeing ‘high risk’ instantly tells me to slow down” (P48). Others emphasized the emotional effect: “The red icon made me stop for a second” (P12) and “It felt safer because I knew the risk level” (P23). These reactions align with prior observations that interpretable cues can convert uncertainty into actionable attention.

Semantic summaries also played a central role in shaping comprehension. Many participants reported that the explanations “finally tell me what the DApp is asking for” (P15), while others noted that the summaries “explained in simple words what would happen after signing” (P10), “told me the purpose clearly” (P11), or “made the action understandable instead of technical” (P36). Several participants connected explanations directly to decision quality: “I could decide if this makes sense instead of hoping it’s safe” (P9) and “It made my decisions feel more deliberate” (P52). These reflections align with prior work showing that clarity is the primary factor that sustains users’ sense of trust~\cite{Lee2004Trust,Abdul2018XAI}.

Transparency of contextual details further strengthened users’ confidence. Participants appreciated seeing wallet addresses, token types, and network information because “everything was shown, including sender and receiver wallets” (P1), “I could check where it came from” (P2), and “the context matched what I expected to see” (P21). Others highlighted the benefit for error detection, noting that “if something looked off, I would notice” (P44). The emphasis on visibility suggests that contextual completeness functions as a verification layer rather than cognitive overhead.

At the same time, participants described challenges related to layout density and information volume. Several found the interface “clear but crowded” (P18), “a lot to process” (P6), or “visually busy when I’m trying to focus” (P45). Others requested finer-grained explanations for moderate-risk judgments: “If it says medium, I want to know why” (P9), “I didn’t know what made it risky” (P11), and “I wanted the reason, not just the label” (P36). A few participants expressed caution about over-reliance: “If it ever got something wrong, that would be worse than no explanation” (P10), underscoring the delicate balance between helpful abstraction and perceived accuracy~\cite{PoursabziSangdeh2021Interpretability}.

Taken together, these reflections show that semantic decoding helped users think before they trusted, enabling comprehension to develop into confidence while preserving their agency. Its value extends beyond usability, shaping how individuals interpret and manage digital risk by providing clarity, contextual grounding, and a sense of control. At the same time, the responses reveal the need to balance helpful structure with information load, and to pair risk labels with sufficient context to avoid over-reliance or misunderstanding. These insights support \textbf{RQ3} and indicate that semantic transparency fosters more thoughtful, cautious, and trustworthy engagement in secure signing.

\section{Discussion}

Our findings show that semantic transparency not only improves accuracy but also reshapes how users perceive and engage with wallet signing. What was once routine and opaque became a moment of reflection and active reasoning. We discuss these implications from three perspectives.

\textbf{Interpretive Security as Human–System Alignment.}
Our findings show that most failures in wallet signing are not cryptographic but epistemic: users do not misverify signatures—they misunderstand them~\cite{Whitten1999Johnny}. This shifts the problem from mathematics to meaning. Cryptographic validity ensures that a message is signed correctly; interpretive validity ensures that the signer understands its intent. The \textit{Signature Semantic Decoder} demonstrates that these forms of validity can coexist through semantic reconstruction that aligns machine logic with human reasoning. By converting raw payloads into contextual explanations, the system reduces the cognitive asymmetry between blockchain protocols and user understanding~\cite{Norman2013Design}, forming what we term \textit{interpretive security}. This view echoes human-centered explainability work in HCI, which holds that intelligibility underpins proper trust calibration rather than serving as an optional veneer~\cite{Abdul2018XAI,Lee2004Trust,Parasuraman1997Humans}. Participants trusted the system because they could inspect and make sense of its interpretation. In this way, interpretability becomes a mechanism for human–system alignment, making security both verifiable and comprehensible.

\textbf{Explanation as Cognitive Mediation.}
Contrary to the common belief that more information inevitably increases cognitive strain, our results show that well-structured explanations can facilitate reasoning rather than overload it. Participants using the semantic interface reported lower mental demand and frustration on NASA TLX scales, even though they spent slightly more time per task. Qualitative feedback indicates that this additional time reflected thoughtful engagement. This finding supports prior evidence that the effectiveness of transparency depends on how explanations are organized rather than on their quantity~\cite{PoursabziSangdeh2021Interpretability}. From a cognitive standpoint, explanations that filter irrelevant details and foreground essential cues can reduce effort while improving judgment~\cite{Sweller1988CLT}. Participant feedback also highlighted a balance between clarity and conciseness: some found the interface visually dense, while others wanted clearer justifications for mid-level risk ratings. These observations point toward \textit{progressive explainability}, where explanation depth adapts to user expertise, context, and task complexity~\cite{Abdul2018XAI}. Users trusted the system because they could inspect and reason about its interpretation; interpretability therefore acts as a human–system alignment mechanism that makes security both verifiable and comprehensible.

\textbf{From User Protection to User Empowerment.} A central implication of this work lies in reframing how security systems relate to users. Traditional wallet design treats users as entities to be protected from mistakes, relying on warnings and constraints that limit autonomy in the name of safety~\cite{Albayati2021WalletUX}. Yet decades of usable-security research show that generic warnings often cause habituation or misinterpretation when context is missing~\cite{Sunshine2009CryingWolf,Akhawe2013Warningland}. Our semantic interface instead views users as partners capable of reasoning when given meaningful cues, aligning with permission and consent research showing that structured, comprehensible disclosures improve decisions over raw permission lists~\cite{Felt2012AndroidPerm,Kelley2012Conundrum}. Participants often described the system as “being on the same side,” marking a shift from protective opacity to empowering transparency~\cite{Binns2018Agency}. Beyond cryptocurrency wallets, the same principle extends to OAuth permissions, AI delegation, and intelligent automation where oversight relies on interpretive visibility~\cite{Abdul2018XAI,Lee2004Trust}. Embedding semantic reasoning at the interface layer moves us from protecting users through opacity to empowering them through understanding.

\section{Limitations and Future Work}
While the controlled experimental setup ensured strong internal validity, it inevitably differed from real-world signing contexts where users face financial stakes, time pressure, and emotional arousal. This abstraction allowed clear isolation of cognitive effects but may limit ecological validity, since participants interacted with simulated interfaces rather than executing live transactions; behaviors such as hesitation, verification, or post-decision regret may therefore not have been fully captured. Future work should extend evaluation to in-situ or incentivized environments that mirror authentic wallet use, where multitasking, mobile constraints, and social context may shape users’ trust calibration and decision pace~\cite{voskobojnikov2021u,voskobojnikov2021u}.
In terms of technical scope, the prototype supports only Ethereum-standard methods (\textit{eth\_sendTransaction}, \textit{personal\_sign}, \textit{eth\_signTypedData}), limiting generalizability to EVM-compatible systems. Extending the framework to multi-chain and account-abstraction architectures~\cite{EIP4337,Zhang2023AA,Zhou2024CrudiTEE,Geels2024Signatures} will test scalability across diverse signing protocols, while integration with smart contract wallets and intent-centric architectures~\cite{AnomaIntent,lashbotsSUAVE} could enable cross-domain signing pipelines.
Another direction lies in adaptive explanation mechanisms that adjust depth to user expertise and transaction complexity~\cite{Abdul2018XAI,PoursabziSangdeh2021Interpretability}, balancing informativeness and cognitive economy. Incorporating real-time threat intelligence or on-chain reputation data~\cite{SecurityAspectsWallets2023,yu2024walletchoice} may strengthen contextual reliability by grounding semantic cues in verifiable evidence. Finally, longitudinal and field studies should examine whether interpretability effects persist over time, whether semantic cues foster lasting signing literacy or short-term caution. As wallets evolve toward more autonomous transaction models that depend on sustained user trust and accountability~\cite{Frohlich2022BlockchainHCI,Elsden2018MakingSense}. Future evaluations could also compare diverse user groups, such as novices and professional traders, to understand how domain expertise shapes the perception and utility of semantic explanations.

\section{Conclusion}

We set out to close the interpretability gap in cryptocurrency wallet signing through empirical inquiry, system design, and controlled evaluation. Two formative studies showed that users often misunderstood signature requests and relied on superficial cues instead of genuine semantic comprehension. Building on these insights, we developed the \textit{Signature Semantic Decoder}, a middleware that reconstructs and presents signing intent through structured parsing and contextual risk cues. A subsequent user study demonstrated that this semantic transparency improved accuracy, clarity, and confidence while lowering cognitive workload, establishing interpretive transparency as a basis for trustworthy and usable cryptographic interaction. Future work will extend this framework to multi-chain and autonomous signing scenarios to ensure that users truly understand what they authorize.


% % use section* for acknowledgment
% \ifCLASSOPTIONcompsoc
%   % The Computer Society usually uses the plural form
%   \section*{Acknowledgments}
% \else
%   % regular IEEE prefers the singular form
% \fi



%\bibliographystyle{IEEEtran}


% \end{thebibliography}

\bibliographystyle{IEEEtran}
\bibliography{reference}

\appendices

\section{Participant Demographics}
\label{Appendix:participant_table}
Table~\ref{tab:appendix_demographics} summarizes the demographic characteristics of participants in the main study (N = 128).

\begin{table}[h]
\vspace{-0.2cm}
\centering
\caption{Participant Demographics}
\label{tab:appendix_demographics}
\renewcommand{\arraystretch}{1.15}
\setlength{\tabcolsep}{4pt}
\begin{tabularx}{\columnwidth}{@{\hspace{0.8em}}X|
>{\raggedright\arraybackslash}p{0.14\columnwidth}|
>{\raggedright\arraybackslash}p{0.14\columnwidth}@{}}
\toprule
\textbf{Characteristics} & \textbf{Number} & \textbf{Percent} \\
\midrule
\multicolumn{3}{c}{\textbf{Gender}} \\
\midrule
Male & 82 & 64.06\% \\
Female & 46 & 35.94\% \\
\midrule
\multicolumn{3}{c}{\textbf{Age Group}} \\
\midrule
18--24 years & 25 & 19.53\% \\
25--34 years & 62 & 48.44\% \\
35--44 years & 23 & 17.97\% \\
45+ years & 18 & 14.06\% \\
\midrule
\multicolumn{3}{c}{\textbf{Crypto Wallet Experience}} \\
\midrule
Less than 6 months & 9 & 7.03\% \\
6--12 months & 22 & 17.19\% \\
1--3 years & 50 & 39.06\% \\
Over 3 years & 47 & 36.72\% \\
\midrule
\multicolumn{3}{c}{\textbf{Main Wallet Used (Multiple selections)}} \\
\midrule
MetaMask & 50 & 39.06\% \\
Phantom & 26 & 20.31\% \\
Coinbase Wallet & 87 & 67.97\% \\
Trust Wallet & 50 & 39.06\% \\
Other & 21 & 16.41\% \\
\midrule
\multicolumn{3}{c}{\textbf{Wallet Type Used (Multiple selections)}} \\
\midrule
Browser extension & 66 & 51.56\% \\
Mobile app & 104 & 81.25\% \\
Hardware wallet & 13 & 10.16\% \\
Other & 2 & 1.56\% \\
\midrule
\multicolumn{3}{c}{\textbf{Education}} \\
\midrule
High School or Belowh & 12 & 9.38\% \\
Undergraduate Student & 11 & 8.59\% \\
Bachelor's Degree & 62 & 48.44\% \\
Graduate / PhD & 42 & 32.81\% \\
\midrule
\multicolumn{3}{c}{\textbf{Occupation}} \\
\midrule
University Student & 10 & 7.81\% \\
Employee & 107 & 83.59\% \\
Self-employed & 7 & 5.47\% \\
Unemployed / Other & 4 & 3.13\% \\
\midrule
\multicolumn{3}{c}{\textbf{IT-related background}} \\
\midrule
Yes & 85 & 66.41\% \\
No & 43 & 33.59\% \\
Prefer not to say & 0 & 0.00\% \\
\bottomrule
\end{tabularx}
\vspace{-0.5cm}
\end{table}

\section{Main Study}
\label{Appendix:main_survey}

This study included a control and an experimental condition. The two groups received identical survey items throughout the study; the only difference lay in the interface shown in the six main tasks. All measurement items, wording, and response formats were held constant to ensure comparability between conditions.

\subsection{Survey}

\noindent\textbf{Demographics:}

\begin{enumerate}[label=\arabic*), leftmargin=2em, labelsep=0.5em, itemsep=0pt, topsep=0pt]
    \item Age range (18--24, 25--34, 35--44, 45+, Prefer not to say)
    \item Gender (Male, Female, Non-binary, Prefer not to say)
    \item Education level (High school or below, Undergraduate, Bachelor’s degree, Master’s degree, Doctorate or above, Other)
    \item Current occupation (Student, Employee, Self-employed, Not currently employed, Other)
    \item IT-related background (Yes—IT-related education or occupation; No; Prefer not to say)
\end{enumerate}

\noindent\textbf{Crypto Experience:}

\begin{enumerate}[label=\arabic*), leftmargin=2em, labelsep=0.5em, itemsep=0pt, topsep=0pt, resume]
    \item How long have you been using crypto wallets? (Less than 6 months, 6--12 months, 1--3 years, Over 3 years)
    \item Which wallet(s) do you mainly use? (MetaMask, Phantom, Rabby, Coinbase Wallet, Trust Wallet, Other)
    \item How often do you encounter wallet signature requests? (Daily, Weekly, Monthly, Rarely)
    \item What type of wallet do you mainly use for signing? (Browser extension, Mobile wallet app, Hardware wallet, Other)
    \item How confident do you feel in understanding what you are signing? (1--5 Likert scale)
    \item Have you ever experienced or suspected a phishing or scam transaction? (Yes, No, Unsure)
    \item What kinds of on-chain activities do you commonly perform? (Token transfers, Token swaps / DEX trading, NFT minting/trading, DeFi lending or staking, Airdrop claiming, Bridging/cross-chain transfer, Other)
\end{enumerate} 
\vspace{0.5em}

\noindent\textbf{Main Task:}

In this task, you will use a browser-based wallet simulator that reproduces the structure and flow of a MetaMask-style confirmation interface. The simulator runs entirely offline, and no real accounts, assets, or blockchain interactions are involved. All signing requests shown to you have been pre-generated from authentic Ethereum contracts representing common scenarios such as token approvals, cross-chain transfers, NFT interactions, and governance actions.

Your task is to carefully examine each signature request and decide whether you would approve or reject it. After each scenario, you will be asked a few short questions about your understanding and perception of what you saw. There are no right or wrong answers; we are interested in your natural reactions and judgment, as if you were making a real decision in your own wallet.

Please review each request attentively before making your decision.

\noindent\textbf{Questions for Each Task:}
\begin{enumerate}[label=\arabic*), leftmargin=2em, labelsep=0.5em, itemsep=0pt, topsep=0pt, resume]
    \item Would you sign this request?(Yes, sign it; No, reject it)
    \item How risky do you think this request is? (1 = Not risky at all; 5 = Extremely risky)
    \item How clear was the meaning of this request to you? (1 = Not clear at all; 5 = Very clear)
    \item How confident are you in your decision above? (1 = Not confident at all; 5 = Very confident)
\end{enumerate}

\noindent\textbf{NASA–TLX Ratings}

\noindent The following items use a continuous 0–100 rating scale.
\begin{enumerate}[label=\arabic*), leftmargin=2em, labelsep=0.5em, itemsep=0pt, topsep=0pt, resume]
    \item Mental Demand: How much mental and perceptual activity was required (e.g., thinking, deciding, calculating, remembering, looking, searching, etc)? 
    \item Physical demand: How much physical activity was required (e.g., pushing, pulling, turning, controlling, activating, etc.)? 
    \item Temporal demand: How much time pressure did you feel due to the rate or pace at which the mission occurred? 
    \item Performance: How successful do you think you were in accomplishing the goals of the mission?
    \item Effort: How hard did you have to work (mentally and physically) to accomplish your level of performance?
    \item Frustration:How discouraged, stressed, irritated, and annoyed versus gratified, relaxed, content, and complacent did you feel during your mission?
\end{enumerate}

\noindent\textbf{Evaluation of Prototype}

\noindent All items below use a 7-point Likert scale (1 = Strongly disagree, 7 = Strongly agree).

\begin{enumerate}[label=\arabic*), leftmargin=2em, labelsep=0.5em, itemsep=1pt, topsep=3pt]

% ---------- TRUST ----------
\item[]\textbf{Trust}
\item I trust the system to make the right decisions.
\item The system behaved reliably during the signing process.
\item I felt confident relying on the system’s interpretation.

% ---------- TRANSPARENCY ----------
\item[]\textbf{Transparency}
\item The meaning of the signature request was clear to me.
\item I understood what would happen after I signed.
\item The interface provided sufficient detail for me to verify the request on my own.

% ---------- SAFETY ----------
\item[]\textbf{Safety}
\item The interface helps me avoid harmful or risky actions.
\item I could easily identify any potential risks in the request.
\item I felt protected from unintended or malicious actions while using the system.

% ---------- USABILITY ----------
\item[]\textbf{Usability}
\item The interface made important factors (amount, allowance, chain, deadline) clear to me.
\item Overall, it was easy to make decisions with this interface.
\item I would rely on this system for similar tasks in the future.

\end{enumerate}

\noindent\textbf{User Feedback}

\begin{enumerate}[label=\arabic*), leftmargin=2em, labelsep=0.5em, itemsep=1pt, topsep=3pt]

\item What aspects of the signing interface did you find most helpful or confusing during the task? (long open text)

\item If you could improve the signing interface for future users, what would you change or add? (long open text)

\end{enumerate}

\subsection{Task Descriptions}

\begin{enumerate}[label=\arabic*), leftmargin=2em, labelsep=0.5em, itemsep=6pt, topsep=4pt]

    \item \textbf{T1 — Opensea Login }  \\
    Log in to OpenSea: You are attempting to log in to the NFT marketplace OpenSea using your crypto wallet. The platform requests you to sign a message to verify ownership of your address.

    \item \textbf{T2 — NFT Mint }  \\
    NFT Minting: You are using a digital art marketplace that allows creators to mint new NFTs directly from their wallet. After uploading your artwork details, the platform prompts a signature request to confirm the creation of your NFT.

    \item \textbf{T3 — DAO Vote }  \\
    Governance Vote: You are visiting a decentralized application related to community governance. After reviewing a proposal, the platform prompts a signature request to record your vote through your wallet.

    \item \textbf{T4 — Bridge / Swap }  \\
    Cross-Chain Transfer: You are using a decentralized application that allows you to move tokens between different blockchain networks. Before the transfer can be processed, the platform asks you to review and confirm a signature request in your wallet.

    \item \textbf{T5 — Unlimited Approval }  \\
    DApp Token Approval: You are visiting a decentralized application that requests permission to use your tokens. A signature window appears, asking you to confirm a request.

    \item \textbf{T6 — Phishing Request }  \\
    Transfer Request: You are interacting with a decentralized application that asks for a transaction confirmation in your wallet. The request appears to involve an Ethereum transfer, and a confirmation window is displayed for your review.

\end{enumerate}

\noindent Please carefully review the request and decide whether to sign it or reject it.

\end{document}